\documentclass[prl,twocolumn,amsmath,amssymb]{revtex4-2}

\usepackage{graphicx}% Include figure files
\usepackage{dcolumn}% Align table columns on decimal point
\usepackage{bm}% bold math
\usepackage{mathrsfs}
\usepackage{subfigure}
\usepackage{natbib}
\usepackage{amsmath}
\usepackage{amssymb}
\usepackage{Jonasmacros}
\usepackage{mathptmx}
\usepackage{txfonts}

\usepackage{pifont}

\usepackage[colorlinks,citecolor=magenta,linkcolor=red]{hyperref}
\usepackage[usenames,dvipsnames,svgnames]{xcolor}

\begin{document}
\date{\today}

\title{Chiral Phonon Induced Spin-Polarization}

%\author{S. Kalh\"ofer}

%\author{M. Shiranzaei}
%\phone{+46 (0)70 167 9264}
\author{J. Fransson}
\email{Jonas.Fransson@physics.uu.se}
\affiliation{Department of Physics and Astronomy, Box 516, 751 20, Uppsala University, Uppsala, Sweden}
%		$^2$Department of Chemical Physics, Weizmann Institute, 76100 Rehovot, Israel}

%\author{L. Nordstr\"om$^1$}
%\affiliation{Department of Physics and Astronomy, Box 516, 75120, Uppsala University, Uppsala, Sweden}

%\abbreviations{IR,NMR,UV}
%\keywords{American Chemical Society, \LaTeX}

%\begin{document}

%\begin{tocentry}
%\begin{center}
%%\includegraphics[width=9cm]{TOC}
%\end{center}
%\end{tocentry}

\begin{abstract}
The current understanding of chirality suggests the existence of a connection between structure and angular momentum, including spin. This is particularly emphasised in the chiral induced spin selectivity effect, where chiral structures act as spin filters. However, the recent discovery of chiral phonons have demonstrated that phonons too may carry angular momentum which also can be regarded as magnetic moments which add to the total moment. Here, it is shown that chiral phonons may induce a non-trivial spin-texture in an otherwise non-magnetic electronic structure. By considering a set-up in which electrons and phonons are interfaced with each other, it is shown that chiral phonons may transfer its angular momentum into the electron reservoir which, thereby, becomes spin-polarized. It is, moreover, shown that an equivalent mechanism does not exist whenever the electrons are interfaced with achiral phonons.
\end{abstract}
\maketitle

%%%%%%%%%%%%%%%%%%%%%%%%%%%%%%%%%%%%%%%%%%%%%%%%%%%%%%%%%%%%%%%%%%%%%%%%%
%%%%%%%%%%%%%%%%%%%%%%%%%%%%%%%%%%%%%%%%%%%%%%%%%%%%%%%%%%%%%%%%%%%%%%%%%
%%%%%%%%%%%%%%%%%%%%%%%%%%%%%%%%%%%%%%%%%%%%%%%%%%%%%%%%%%%%%%%%%%%%%%%%%
%%%%%%%%%%%%%%%%%%%%%%%%%%%%%%%%%%%%%%%%%%%%%%%%%%%%%%%%%%%%%%%%%%%%%%%%%
%%%%%%%%%%%%%%%%%%%%%%%%%%%%%%%%%%%%%%%%%%%%%%%%%%%%%%%%%%%%%%%%%%%%%%%%%
%%%%%%%%%%%%%%%%%%%%%%%%%%%%%%%%%%%%%%%%%%%%%%%%%%%%%%%%%%%%%%%%%%%%%%%%%

%%%%%%%%%%%%%%%%%%%%%%%%%%%%%%%%%%%%%%%%%%%%%%%%%%%%%%%%%%%%%%%%%%%%%%%%%
%%%%%%%%%%%%%%%%%%%%%%%%%%%%%%%%%%%%%%%%%%%%%%%%%%%%%%%%%%%%%%%%%%%%%%%%%
%%%%%%%%%%%%%%%%%%%%%%%%%%%%%%%%%%%%%%%%%%%%%%%%%%%%%%%%%%%%%%%%%%%%%%%%%
%%%%%%%%%%%%%%%%%%%%%%%%%%%%%%%%%%%%%%%%%%%%%%%%%%%%%%%%%%%%%%%%%%%%%%%%%
%%%%%%%%%%%%%%%%%%%%%%%%%%%%%%%%%%%%%%%%%%%%%%%%%%%%%%%%%%%%%%%%%%%%%%%%%
%%%%%%%%%%%%%%%%%%%%%%%%%%%%%%%%%%%%%%%%%%%%%%%%%%%%%%%%%%%%%%%%%%%%%%%%%

Phonons represent the collective nuclear motion within a structure. As such, phonons are the quantum mechanically defined quantities with which the mechanical degrees of freedom are effectively incorporated into the framework of the general quantum field theory. While traditionally being regarded as a quantity carrying linear momentum, it is only recently that angular momentum of phonons have been considered. An incomplete list of important results are phonon Hall, phonon spin Hall, and phonon angular momentum Hall effects \cite{PhysRevLett.95.155901,PhysRevLett.96.155901,PhysRevB.86.104305,PhysRevLett.100.145902,PhysRevLett.105.225901,PhysRevLett.113.265901,NanoLett.20.7694}, phonon contribution to spin-relaxation processes \cite{PhysRevB.92.024421,PhysRevLett.121.027202,PhysRevB.97.174403,PhysRevLett.126.225703} and the Einstein-de Haas effect \cite{PhysRevLett.112.085503,PhysRevB.97.174403,PhysRevB.99.064428,Nature.565.209}, phononically mediated spin-spin interactions \cite{PhysRevMaterials.1.074404,PhysRevLett.110.156402}, temperature gradient induced phonon angular momentum \cite{PhysRevLett.121.175301}, and optically activated chiral phono-magnetic effects \cite{PhysRevMaterials.3.064405,PhysRevResearch.2.043035,PhysRevResearch.4.013129}. Experimentally, progress has been made in observations of chiral phonons \cite{Science.359.579,AdvMater.33.2101618,SciAdv.8.eabm4005,NaturePhys.15.221,PhysRevLett.128.075901}, and phonon induced magneto-thermal properties \cite{PhysRevLett.106.186601,SolidStateComm.283.37,AdvFunctMater.29.1904734}.

The existence of phonon angular momentum opens up the possibility to couple the electronic spin-degrees of freedom with the mechanical. It is well-established that spin and nuclear motion are coupled directly through, e.g., the Elliot-Yafet mechanism \cite{PhysRev.89.689,PhysRev.96.266,JETP.35.846,Yafet1963,NJP.18.023012,PhysRevB.102.235416}, but also indirectly via the electronic structure \cite{PhysRevMaterials.1.074404}, and it has been demonstrated that such coupling opens for a viable explanation of the chiral induced spin selectivity effect \cite{PhysRevB.102.035431,PhysRevB.102.214303,PhysRevB.102.235416,NanoLett.21.3026}.

Hitherto, however, the angular momentum of phonons and electrons have been considered as separate from one another, where the magnetic moment associated with the chiral phonons have been studied in its own right. While this is definitely pertinent, the effects angular momentum transfer between phonons and electrons has been discussed in a semi-classical model \cite{arXiv.2105.08485}, in which a spin-dependent coupling between phonons and electrons is assumed and it is demonstrated that chiral phonons may give rise to the spin-Seebck effect, something that was also recently observed in experiments \cite{JunLiu}. Nonetheless, the mechanism that enables the angular momentum transfer is yet to be discussed. The purpose with this Letter is to present a coherent theory that ties the existence of phononic angular momentum to a broken electronic spin-degeneracy. It is shown that the mechanism is provided through a vibronically assisted spin-orbit interaction and while this coupling is always present, chiral phonons are required for inducing a spin-polarization.

Chirality is a geometrical property where the structure lacks both inversion and reflection symmetries. It is easy to demonstrate that chiral phonons must carry a non-vanishing angular momentum $\bfJ_\text{ph}$.
This can be seen directly from the definition
\begin{align}
\bfJ_\text{ph}(t)=&
	\int
		\bfQ(\bfr,t)\times\dot\bfQ(\bfr,t)
	d\bfr
	,
\end{align}
where $\bfQ$ is the nuclear displacement, which is connected to phonons through the relation $\bfQ(\bfr,t)=\sum_ql_q\bfepsilon_qQ_p(t)e^{i\bfq\cdot\bfr}$, where $Q_q=b_{\bfq\mu}+b^\dagger_{\bar\bfq\mu}$, is the quantum phonon displacement operator. Here, $b_q$ and $b^\dagger_q$ denote the phonon destruction and creation operators, respectively, with $q=(\bfq,\mu)$ ($\bar{q}=(\bar\bfq,\mu)=(-\bfq,\mu)$) comprising the wave vector $\bfq$ and normal mode $\mu$, and $l_q=\sqrt{\hslash/2\rho v\omega_q}$ which defines a length scale in terms of the phonon energy $\omega_q$, density $\rho$, and volume $v$, whereas $\bfepsilon_q$ is the displacement polarization vector.
% satifying $\bfepsilon_{\bfq\mu}^*=\bfepsilon_{\bar\bfq\mu}$.
%, in general, while the more restricted equality $\bfepsilon_{\bfp\mu}^*=\bfepsilon_{\bfp\mu}$ only holds in inversion invariant structures.

%The mechanical angular momentum $\bfJ_\text{ph}$ is defined through the expression

The expectation value of the angular momentum, hence, assumes the form
\begin{align}
\av{\bfJ_\text{ph}}(t)=&
	\lim_{t'\rightarrow t}
	i\dt\sum_{pq}
		l_pl_q
		\bfepsilon_p\times\bfepsilon_{\bar{q}}
		D^>_{pq}(t',t)
		\int
			e^{i(\bfp-\bfq)\cdot\bfr}
		d\bfr
	,
\label{eq-avJph}
\end{align}
where $D^>_{pq}(t',t)=(-i)\av{Q_p(t')Q_{\bar{q}}(t)}$ defines the correlations between the phonons $Q_p(t')$ and $Q_{\bar{p}}(t)$.
The expression in Eq. (\ref{eq-avJph}) shows that a finite phonon angular momentum requires non-collinear polarisations $\bfepsilon_p$ and $\bfepsilon_{\bar{q}}$.

Non-collinear polarizations can on the one hand be achieved when there is a mechanism that mixes the phonon modes $p$ and $q$. Such mode mixing may originate from, e.g., anharmonic effects or scattering off defects acting upon the otherwise orthogonal modes defined in the harmonic approximation.

In the harmonic approximation, the phonons can, up to a constant, be summarised the Hamiltonian form as $\Hamil_\text{ph}=\sum_q\omega_qb^\dagger_qb_q$. In this form, it is assumed that the phonon modes do not mix which, therefore, leads to a vanishing angular momentum.
However, the introduction of a component of the kind $W_{\bfq\mu\nu}b^\dagger_{\bfq\mu}b_{\bfq\nu}$ provides a mode mixing, which can be understood as chirality.

To see this, consider the phonon spinor $\Phi_\bfq=\{b_{\bfq\mu_i}\}_{i=1}^N$, for $N$ modes, which enables us to write the phonon model as
\begin{align}
\Hamil_\text{ph}=&
	\sum_\bfq \Phi^\dagger_\bfq\bfomega_\bfq\Phi_\bfq
	.
\end{align}
Written like this, the phonon spectrum is defined through the matrix $\bfomega_\bfq=\omega_{0\bfq}\tau^0+\bfomega_{1\bfq}\cdot\bftau$, where $\omega_0$ and $\bfomega_1$ represent the mode conservative and mode mixing components, respectively, whereas $\tau^0$ and $\bftau$ are the $N$-dimensional identity and vector of spin matrices. While in this model it is assumed that the mixing only takes place between modes $\mu_i$ and $\mu_j$ with the same momentum $\bfq$, it is straight forward to generalise the model to also include mixing between different momenta.

Considering a structure with two modes, such that $\bftau$ reduces to the Pauli matrices, the phonon Green function $\bfD_\bfq(z)=\av{\inner{\Phi_\bfq}{\Phi_{\bar\bfq}}}(z)$, can be written as
\begin{subequations}
\label{eq-D}
\begin{align}
\bfD_\bfp(z)=&
	2\bfomega_\bfq
	\frac{z^2-\omega_{0\bfq}^2-\omega_{1\bfq}^2+2\omega_{0\bfq}\bfomega_{1\bfq}\cdot\bftau}
	{(z^2-\omega_{0\bfq}^2-\omega_{1\bfq}^2)^2-4\omega_{0\bfq}^2\omega_{1\bfq}^2}
\label{eq-D1}
\\=&
	\frac{1}{2}
	\sum_{s=\pm1}
			\frac{2\omega_{\bfq s}}{z^2-\omega_{\bfq s}^2}
			\Bigl(
				\tau^0+s\hat\bfomega_{1\bfq}\cdot\bftau
			\Bigr)
	%\biggl\{
	%	\frac{\tau^0+\hat\bfomega_{1\bfp}\cdot\bftau}{z^2-(\omega_{0\bfp}+\omega_{1\bfp})^2}
	%	+
	%	\frac{\tau^0-\hat\bfomega_{1\bfp}\cdot\bftau}{z^2-(\omega_{0\bfp}-\omega_{1\bfp})^2}
	%\biggr\}
	,
\label{eq-D2}
\end{align}
\end{subequations}
where $\omega_{\bfq s}=\omega_{0\bfq}+s\omega_{1\bfq}$, $\omega_{1\bfq}=|\bfomega_{1\bfq}|$ and $\hat\bfomega_{1\bfq}=\bfomega_{1\bfq}/\omega_{1\bfq}$. The form of this propagator written in Eq. (\ref{eq-D2}) explicitly describes two modes with opposite helicity, or, chirality.

%It is important to notice that the off-diagonal components of this propagator carry phase factors $e^{\pm i\phi_p}$, where $\phi_p$ defines the azimuthal angle of the momentum vector $\bfp$. Furthermore, it can be concluded that phonons with a non-vanishing mixing component $\bfomega_{1\bfp}$ necessarily carry angular momentum. This angular momentum can, moreover, be transferred from the phonon subsystem to the electronic and, thereby, induce an electronic magnetic moment. This assertion is justified in the following discussion.

On the other hand, the polarization for a chiral mode is neither reflection nor inversion symmetric, where the latter condition leads to that $\bfepsilon^*_{\bfq\mu}=\bfepsilon_{\bar\bfq\mu}\neq\bfepsilon_{\bfq\mu}$, since equality in the last relation requires inversion symmetry. For, e.g., a helical structure with transversal and longitudinal lattice parameters $a$ and $c$, respectively, the polarization may be expressed as $\bfepsilon_\bfq=(a\cos\phi_q,a\sin\phi_q,\mathcent\phi_q)/d(\phi_q)$, where $\mathcent=c/2\pi$ and $d(\phi_q)=\sqrt{a^2+\mathcent^2\phi_q^2}$, which displays a variation of the mode that depends on the azimuthal angle $\phi_q$. Hence, for such a mode, the vector product $\bfepsilon_\bfq\times\bfepsilon_{\bar\bfq}=(-a\mathcent\phi_q\sin\phi_q,a\mathcent\phi_q\cos\phi_q,-a^2\sin2\phi_q)/d^2(\phi_q)$, which suggests that a free chiral phonon mode, for which $D^>_\bfq(t,t')=(-i)[n_B(\omega_\bfq)e^{-i\omega_\bfq(t-t')}-n_B(-\omega_\bfq)e^{i\omega_\bfq(t-t')}]$, where $n_B(\omega)$ is the Bose-Einstein distribution function, carry the non-vanishing angular momentum
\begin{align}
\av{\bfJ_\text{ph}}=&
	\frac{\omega_\bfq l_\bfq^2}{\pi d^2(\phi_q)}
	\begin{pmatrix}
		-a\mathcent\phi_q\sin\phi_q \\
		a\mathcent\phi_q\cos\phi_q \\
		-a^2\pi\sin2\phi_q
	\end{pmatrix}
	.
\end{align}

It is important to notice that the off-diagonal components of the phonon propagator, Eq. \eqref{eq-D2}, carry the phase factors $e^{\pm i\phi_p}$, where $\phi_p$ defines the azimuthal angle of the momentum vector $\bfp$. This phase dependence reoccurs also in the helical polarization vector and is an essential feature of chirality. Due to this phase dependence, the phonons necessarily carry angular momentum.
%%%%%%%%%%%%%%%%%%%%%%%%%%%%%%%%%%%%%%%%%%%%%%%%%%%%%%%%%%%%%%%%%%%%%%%%%
%%%%%%%%%%%%%%%%%%%%%%%%%%%%%%%%%%%%%%%%%%%%%%%%%%%%%%%%%%%%%%%%%%%%%%%%%
%%%%%%%%%%%%%%%%%%%%%%%%%%%%%%%%%%%%%%%%%%%%%%%%%%%%%%%%%%%%%%%%%%%%%%%%%
%%%%%%%%%%%%%%%%%%%%%%%%%%%%%%%%%%%%%%%%%%%%%%%%%%%%%%%%%%%%%%%%%%%%%%%%%
%%%%%%%%%%%%%%%%%%%%%%%%%%%%%%%%%%%%%%%%%%%%%%%%%%%%%%%%%%%%%%%%%%%%%%%%%
%%%%%%%%%%%%%%%%%%%%%%%%%%%%%%%%%%%%%%%%%%%%%%%%%%%%%%%%%%%%%%%%%%%%%%%%%

The objective in this Letter, is to show that the phononic angular momentum can be transferred to the electronic subsystem and, hence, induce a spin-polarization. For this sake, consider the coupling between electrons and phonons, which generally can be written as \cite{PhysRevB.102.235416}
\begin{align}
\Hamil_\text{e-ph}=&
	\sum_{\bfk q}
		\psi^\dagger_{\bfp+\bfk}\bfU_{\bfk q}\psi_\bfk Q_q
	,
\end{align}
where the coupling matrix $\bfU_{\bfk q}=U_{\bfk q}\sigma^0+\bfJ_{\bfk q}\cdot\bfsigma$ accounts for a spin-conservative electron-phonon coupling, $U_{\bfk q}$, and an electron-phonon assisted spin-orbit interaction, $\bfJ_{\bfk q}$.

%, e.g., $\Hamil_\text{el}=\sum_\bfk\dote{\bfk}\psi_\bfk^\dagger\psi_\bfk$, where $\dote{\bfk}$ defines the electronic dispersion relation.
The spin-polarisation $\av{\bfM_\bfk}$ of the electrons are calculated using the identity $\av{\bfM_\bfk}=(-i){\rm sp}\bfsigma\int\bfG_{\bfk\bfk}^<(\omega)d\omega/4\pi$, where $\bfG^<_{\bfk\bfk'}(\omega)$ defines the lesser form of the general single electron Green function $\bfG_{\bfk\bfk'}(z)=\av{\inner{\psi_\bfk}{\psi_{\bfk'}^\dagger}}(z)$, whereas ${\rm sp}$ denotes the trace over spin 1/2 space. To the second order (Hartree-Fock) approximation in the electron-phonon coupling, this Green function can be calculated from the Dyson equation
\begin{subequations}
\begin{align}
\bfG_{\bfk\bfk'}(z)=&
	\delta(\bfk-\bfk')\bfg_\bfk(z)
	+
	\bfg_\bfk(z)\sum_{\bfkappa}\bfSigma^\text{(HF)}_{\bfk\bfkappa}(z)\bfG_{\bfkappa\bfk'}(z)
	,
\\
\bfSigma^\text{(HF)}_{\bfk\bfk'}(z)=&
	(-i)\delta(\bfk-\bfk')
	\sum_{qq'}
		D_{qq'}
		\bfU_{\bfk q}
		{\rm sp}
			\int\bfG^<_{\bfk\bfk}(\omega)\bfU_{\bfk\bar{q}'}\frac{d\omega}{2\pi}
\nonumber\\&
	-
	\frac{1}{\beta}
	\sum_{z_\nu qq'}
		\bfU_{\bfk q}\bfG_{\bfk\bfk'}(z+z_\nu)\bfU_{\bfk'\bar{q}'}D_{qq'}(z_\nu)
	,
\end{align}
\end{subequations}
where $D_{qq'}=\int\delta(\omega)D^a_{qq'}(\omega)d\omega$.

While this equation should be self-consistently solved, for an analysis of the induced spin symmetries, it is sufficient to replace the electronic Green functions in these expressions with its unperturbed form $\bfg_\bfk(z)=\sigma^0g_\bfk(z)$, where $g_\bfk(z)=1/(z-\dote{\bfk})$ and $\bfg^<_\bfk(\omega)=2\pi\sigma^0f(\omega)\delta(\omega-\dote{\bfk})$, whereas $f(\omega)$ is the Fermi-Dirac distribution function. These replacements lead to the simplified self-energy
\begin{align}
\bfSigma_\bfk^\text{(HF)}(z)=&
	2f(\dote{\bfk})
	\sum_{qq'}
		\delta(\bfq)
		U_{\bfk q}
		D_{qq'}
		\bfU_{\bfk\bar{q}'}
\nonumber\\&
	-
	\frac{1}{\beta}
	\sum_{\nu qq'}
		g_{\bfk-\bfq}(z-z_\nu)
		\bfU_{\bfk q}
		D_{qq'}(z_\nu)
		\bfU_{\bfk\bar{q}'}
	.
\label{eq-SigmaHFsimp}
\end{align}
The Hartree (first) contribution is to lowest order linear in the component $\bfJ_{\bfk q}\cdot\bfsigma$ which, hence, shows that the coupling between the electrons and phonons may break time-reversal symmetry. However, this contribution vanishes in this approximation since there are no phonons at $\bfp=0$. By contrast, the exchange (second) contribution does not only open for breaking the time-reversal symmetry, since in general
\begin{align}
\bfU_{\bfk p}\bfU_{\bfk q}=&
	U_{\bfk q}U_{\bfk q'}
	+
	\bfJ_{\bfk q}\cdot\bfJ_{\bfk q'}
\nonumber\\&
	+
	\Big(
		U_{\bfk q}\bfJ_{\bfk q'}
		+
		\bfJ_{\bfk q}U_{\bfk q'}
		+
		i\bfJ_{\bfk q}\times\bfJ_{\bfk q'}
	\Bigr)
	\cdot\bfsigma
	,
\end{align}
but it also provides a correlation between the electronic and phononic degrees of freedom.

However, before showing that chiral phonon leads to a breaking of the electronic spin-degeneracy, it is pertinent to discuss the origin of the phonon induced spin-polarization and spin-flip processes which are enabled by $\bfJ_{\bfk q}$. For this purpose, it is instructive to consider the spin-orbit coupling, which can be written as \cite{PhysRevB.91.174415}
\begin{align}
H_\text{SOC}=&
	\frac{\xi}{2}
	\Bigl[
		\bfE\times\bfp
		-
		\bfp\times\bfE
	\Bigr]
	\cdot
	\bfsigma
	,
\end{align}
where $\bfE$ is the total electric field acting on the electrons, whereas $\xi=1/4c^2$ in atomic units. By introducing the field operator $\psi(\bfr)=\int\psi_\bfk e^{i\bfk\cdot\bfr}d\bfk/\Omega$ and its Hermitian conjugate, and assuming time-independent magnetic fields $\bfB$, such that $\nabla\times\bfE=-\dt\bfB=0$, the spin-orbit coupling contribution can in the second quantization be written
\begin{align}
\Hamil_\text{SOC}=&
	\calE_S
	-
	i\xi
	\int
		\psi_\bfk^\dagger
		(\bfk\times	\bfk')
		\cdot
		\bfsigma
		V(\bfr)
		e^{-i(\bfk-\bfk')\cdot\bfr}
		\psi_{\bfk'}
	d\bfr
	\frac{d\bfk}{\Omega}
	\frac{d\bfk'}{\Omega}
	,
\label{eq-HSOC}
\end{align}
where $\calE_S$ is the boundary surface contribution, and where the electric field has been identified with a the electronic confinement potential $V(\bfr$) using the relation $\bfE=-\nabla V(\bfr)$.

Here, $\calE_S=\int\psi_\bfk\calS_{\bfk\bfk'}\psi_{\bfk'}d\bfk d\bfk'/\Omega^2$, where $\calS_{\bfk\bfk'}$ is given by
\begin{align}
\calS_{\bfk\bfk'}=&
	\xi
	\int_\calS
		(\bfk\times\hat\bfn)
		\cdot
		\bfsigma
		V(\bfr)e^{-i(\bfk-\bfk')\cdot\bfr}
	d\calS
	,
\end{align}
where $\hat\bfn$ denotes the outward normal on the boundary surface $\calS$. This contribution can be identified as the Rashba spin-orbit coupling.

The confinement potential $V(\bfr)=\int V(\bfr-\bfR)d\bfR$, where $\bfR=\bfR_0+\bfQ(\bfR)$ denotes the coordinate for the nuclei in terms of the equilibrium position $\bfR_0$ and displacement $\bfQ(\bfR)$. For small displacements, the potential can be expanded around the equilibrium positions $V(\bfr-\bfR)\approx V(\bfr-\bfR_0)-\bfQ\cdot[\nabla_\bfR V(\bfr-\bfR)]_{\bfR\rightarrow\bfR_0}$, where the first contribution along with the surface integral provide the static spin-orbit interaction.

The spin-orbit interaction with the lowest order coupling to the nuclear displacement is given by the Hamiltonian
\begin{align}
\Hamil^{(1)}_\text{SOC}=&
	(-i)\hslash\xi
	\int
		\psi^\dagger_\bfk
		\Bigl[
			(\bfk\times\bfk')\cdot\bfsigma
		\Bigr]
		\Bigl[
			\bfQ\cdot[\nabla_\bfR V(\bfr-\bfR)]_{\bfR\rightarrow\bfR_0}
		\Bigr]
\nonumber\\&\times
		e^{-i(\bfk-\bfk')\cdot\bfr}
		\psi_{\bfk'}
	d\bfR_0
	d\bfr
	\frac{d\bfk}{\Omega}
	\frac{d\bfk'}{\Omega}
	.
\end{align}
Using $\nabla_\bfR V(\bfr-\bfR)=(-i)\int\bfq V(\bfq)e^{i\bfq\cdot(\bfr-\bfR)}d\bfq/\Omega$, leads to that the phonon assisted spin-orbit interaction can be written
\begin{align}
\Hamil^{(1)}_\text{SOC}=&
	\int
		\psi^\dagger_{\bfk+\bfq}
			\bfJ_{\bfk q}\cdot\bfsigma
			Q_q
		\psi_\bfk
	\frac{d\bfq}{\Omega}
	\frac{d\bfk}{\Omega}
	,
\end{align}
where $\bfJ_{\bfk q}=i\xi U_{\bfk q}(\bfk\times\bfq)$, where $U_{\bfk q}=U_q=il_q\bfepsilon_q\cdot\bfq V(\bfq)$ is the spin-conservative electron-phonon interaction. This gives the total electron-phonon coupling $\bfU_{\bfk q}=U_q\sigma^0+\bfJ_{\bfk q}\cdot\bfsigma$.

From an order of magnitude estimation, it can be deduced that the ratio exchange diagram originating from the spin-dependent and spin-conservative processes, that is, those which are proportional to $U_{\bfk q}\bfJ_{\bfk\bar{q}}$ and $U_{\bfk q}U_{\bfk\bar{q}}$, respectively, is roughly $2\xi q_c/3$ ($\times\bfk$), where $q_c$ is the momentum corresponding to the phononic high energy cut-off.

More importantly, however, is that while $J_{\bfk q}$ is linear in the spin-orbit coupling parameter $\xi$, that is, in the same order as the static spin-orbit interaction contribution $\Hamil_\text{SOC}^{(0)}$, the phonon assisted contribution may potentially have a deeply profound influence on the electronic structure. The reason is that the coupling to phonons by all means generates many-body effects, that is, electronic correlations, which may lead to spontaneous symmetry breaking of the electrons.

To see this, consider the influence from the chiral phonons on the magnetic moment $\av{\bfM}=\int\av{\bfM_\bfk}d\bfk/\Omega$. In the first order approximation of the lesser Green function, $\bfG^<_\bfk\approx\bfg^<_\bfk+\bfg^r_\bfk\bfSigma^{\text{(HF)},<}_\bfk\bfg^a_\bfk$, the $\bfk$-resolved spin-moment reduces to the expression ($\Omega=(2\pi)^3$)
\begin{subequations}
\begin{align}
\av{\bfM_\bfk}=&
	4\xi
	\int
		f(\dote{\bfk})
		l_q^2
		V^2(\bfq)
		\bfK_{\bfk\bfq}
	\frac{d\bfq}{(2\pi)^3}
	,
\label{eq-Mk}
\\
\bfK_{\bfk\bfq}=&
	|\bfepsilon_q\cdot\bfq|^2
	(\bfk\times\bfq)
	\sum_{ss'=\pm1}
		s'n_B(s'\omega_{\bfq s})
		\frac{1+s\hat\omega_{\perp\bfq}\cos\phi_q}{(\dote{\bfq}+s'\omega_{\bfq s}-\bfk\cdot\bfq)^2}
	,
\label{eq-K}
\end{align}
\end{subequations}
where $\hat\omega_{\perp\bfq}=|(\omega_{1\bfq}^{(x)},\omega_{1\bfq}^{(y)},0)|/\omega_{1\bfq}$.

\begin{figure}[t]
\begin{center}
\includegraphics[width=\columnwidth]{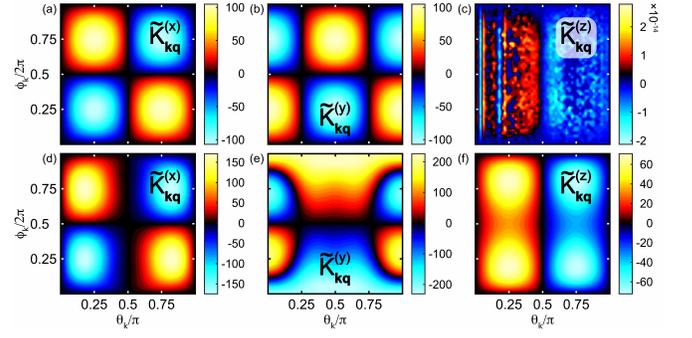}%{tildeKkqxyz.jpg}%
\end{center}
\caption{
Examples of $\tilde{\bfK}_{\bfk\bfq}=(\tilde{K}_{\bfk\bfq}^{(x)},\tilde{K}_{\bfk\bfq}^{(y)},\tilde{K}_{\bfk\bfq}^{(z)})$ as function of the polar (horizontal) and azimuthal (vertical) angles $\theta_k$ and $\phi_k$, for (a)--(c) achiral and (d)--(f) chiral phonons. Parameters used are (a.u.) $k=\sqrt{2m_e\dote{}}/\hslash$, $\dote{}=0.15$ eV, $\omega_{\bfq s}=(\omega_0+s\omega_1)q$, $\omega_0q=0.04$ eV, $\omega_0=1$ , $\omega_1=\omega_0/10$, and $\bfepsilon=(0,0,1)$ at $T=300$ K.
}
\label{fig-tildeK}
\end{figure}

By integrating out the phases $\phi_q$ and $\theta_q$, defining the spin-texture $\tilde{\bfK}_{\bfk\bfq}=\int_0^\pi\int_0^{2\pi}\bfK_{\bfk\bfq}d\phi_q\sin\theta_qd\theta_q/(2\pi)^2$, under the assumption that $V(\bfq)=V(|\bfq|)$, the resulting kernel is a regular function of the electronic momentum $\bfk$. This is shown in Fig. \ref{fig-tildeK}, where $\tilde{\bfK}_{\bfk\bfq}$ is plotted as function of the polar and azimuthal angles $\theta_k$ and $\phi_k$, respectively, for (a)--(c) achiral and (d)--(f) chiral phonons.

Both for achiral and chiral phonons, there is a vibrationally assisted spin-orbit coupling induced in the electronic structure, which follows on general grounds and has been discussed in different context, e.g., Ref. \cite{PhysRevLett.110.026802}. The absence of a $z$-component for achiral phonons in the example given in Fig. \ref{fig-tildeK} can be understood through the following observation.

The denominator in $\bfK_{\bfk\bfq}$, see Eq. \eqref{eq-K}, depends on the difference between the phononic and electronic phases $\phi_q$ and $\phi_k$, respectively, as $[A_{\bfk\bfq ss'}-B_{\bfk\bfq}\cos(\phi_q-\phi_k)]^2$, where $A_{\bfk\bfq ss'}=q^2/2+s'\omega_{\bfq s}-kp\cos\theta_k\cos\theta_q$ and $B_{\bfk\bfq}=kp\sin\theta_k\sin\theta_q$. This is an even function of the phase difference $\phi_q-\phi_k$ on the interval $[0,2\pi]$, while both $|\bfepsilon_q\cdot\bfq|^2$ and $\bfk\times\bfq$ can be written as sum of even and odd functions of $\phi_q-\phi_k$. For a the polarization $\bfepsilon_q=(0,0,1)$, the numerator of the longitudinal spin-texture $K_{\bfk\bfq}^{(z)}$ is proportional to $\sin(\phi_q-\phi_k)$, which is odd on the interval $[0,2\pi]$. Hence, $\int_0^{2\pi}K_{\bfk\bfq}^{(z)}d\phi_q/2\pi=0$. For chiral phonons, $\hat\omega_{\perp\bfq}\neq0$, which leads to that the numerator of $K_{\bfk\bfq}^{(z)}$ also contains a term which is proportional to $\sin(\phi_q-\phi_k)\cos\phi_q=\sin(\phi_q-\phi_k)[\cos\phi_k\sin(\phi_q-\phi_k)+\sin\phi_k\cos(\phi_q-\phi_k)]$, where the first (second) contribution is an even (odd) function of $\phi_q-\phi_k$. Therefore, $\int_0^{2\pi}K_{\bfk\bfq}^{(z)}d\phi_q/2\pi\neq0$ for chiral phonons.

The above discussion pertains to longitudinally polarized phonons. For transversely ($xy$-plane) polarized phonons, the vibrationally assisted spin-orbit coupling generates a non-vanishing spin-texture for all $\tilde{K}_{\bfk\bfq}^{(i)}$, $i=x,y,z$. However, the longitudinal spin-texture $\tilde{K}_{\bfk\bfq}^{(z)}$ induced for achiral phonons polarized along some transverse axis, e.g., $\hat\bfx$, exactly cancels the spin-texture induced for phonons polarized along its orthogonal transverse axis, e.g., $\hat\bfy$. Chiral phonons, on the other hand, breaks this symmetry such that there remains a net longitudinal spin-texture after summing over all phonon modes.

The conclusion is, that a vibrationally assisted spin-orbit coupling exists for any type of phonons which, in turn, gives rise to a transversal spin-texture. However, a net longitudinal spin-texture only arises in presence of chiral phonons. Nevertheless, as is illustrated in Fig. \ref{fig-tildeK}, no magnetic moment is generated from this spin-texture since its angular dependencies integrate to zero.

\begin{figure}[t]
\begin{center}
\includegraphics[width=\columnwidth]{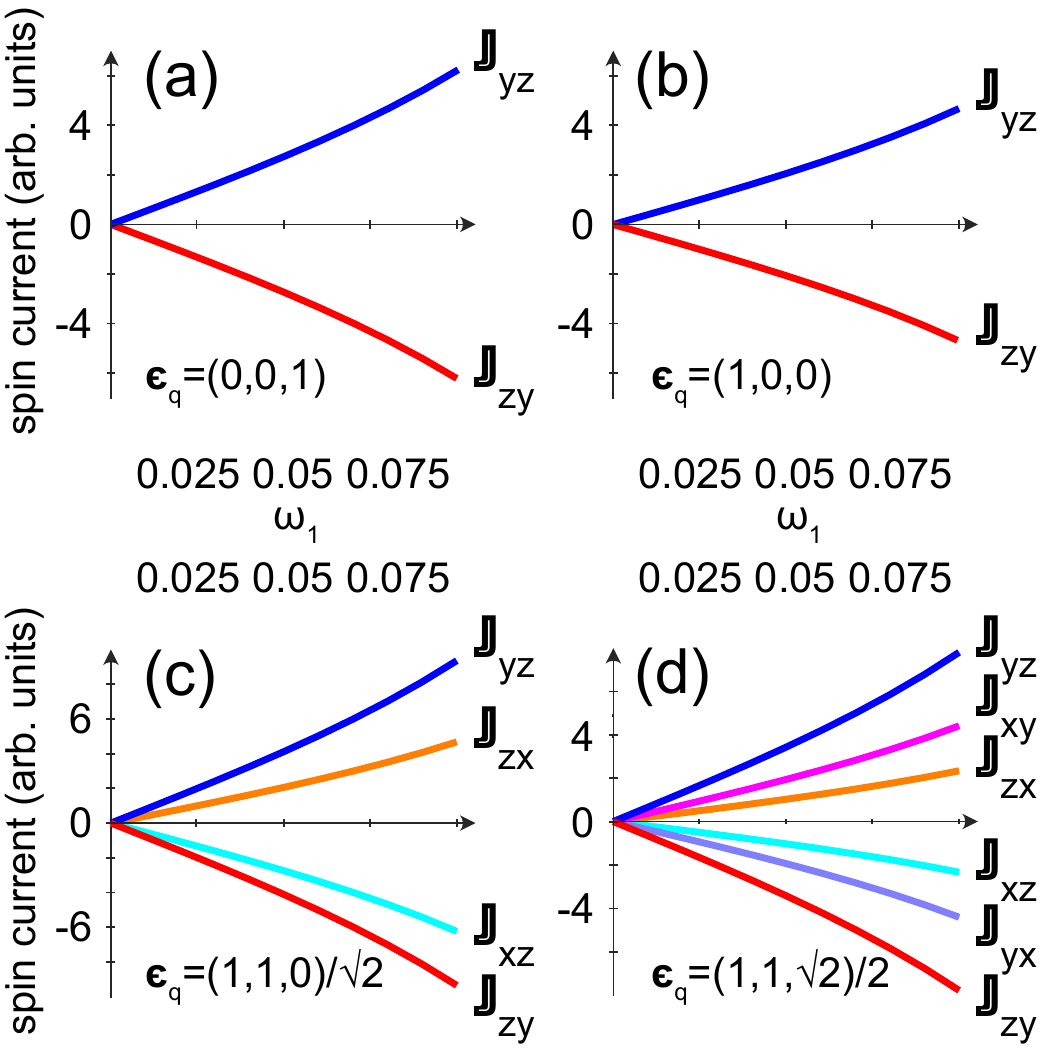}%{tildeKkqxyz.jpg}%
\end{center}
\caption{
Non-zero components of the spin-current tensor as function of the mixing $\omega_1$ for phonons polarized along (a) $\bfepsilon_q=(0,0,1)$, (b) $\bfepsilon_q=(1,0,0)$, (c) $\bfepsilon_q=(1,1,0)/\sqrt{2}$, and (d) $\bfepsilon_q=(1,1,\sqrt{2})/2$. Other parameters are as in Fig. \ref{fig-tildeK}.
}
\label{fig-Jtens}
\end{figure}

A manifestation of the chiral phonon induced spin-texture is the presence of spin-currents. The spin-current can, in the current approximation, be written as $\mathbb{J}(\bfr)=(-i){\rm sp}\bfsigma\int\bfk\bfG^<_\bfk(\omega)d\bfk d\omega/4\pi\Omega=\int\bfk\av{\bfM_\bfk}d\bfk/\Omega$, where the last expression vividly suggests a drift of the spin-texture. This drift is a consequence of the spin-orbit coupling and exists under general conditions, including equilibrium. However, a net spin flow can only be induced by chiral phonons, not by achiral phonon. This is illustrated in Fig. \ref{fig-Jtens}, where the non-zero components of the spin-current tensor are plotted as function of the mode mixing parameter $\omega_1$ for different phonon polarizations. Achiral phonons are represented by $\omega_1=0$, at which condition there is no induced spin-current, regardless of the phonon polarization.

By contrast, chiral phonons induces spin-currents for which the spatial distribution depends on the polarization. As can be seen in Fig. \ref{fig-Jtens}, for phonons polarized solely in the longitudinal or the transverse directions, the spin-currents flow between these directions only. For phonons polarized along some direction with both longitudinal and transverse components, the spin-currents flow between all these. However, in general, the diagonal spin-currents, $\mathbb{J}_{ii}$, $i=x,y,z$, are negligible which is expected for a non-collinear spin-texture.

The induced spin-currents would be measurable as a spin accumulation at, e.g, a terminating surface, by exposing the system to an accelerating force. For instance, a spatially weakly inhomogenous temperature, such that $T=T(\bfr)\approx T_0+(\bfr-\bfr_0)\cdot\nabla_\bfr T(\bfr)_{|\bfr_0}$, where $T_0=T(\bfr_0)$ represents the reference temperature, provides an accelerating field acting on the electrons via the temperature gradient $\nabla T$. Then, to linear order in this field, the longitudinal spin-current partitions into two contributions. The first of these derive from the electronic temperature variations $k_B\beta^2([\dote{\bfk}-\dote{F}]/4)\cosh^{-2}(\beta[\dote{\bfk}-\dote{F}]/2)(\bfr-\bfr_0)\cdot\nabla T$. This contribution gives the expected drift of electrons around the Fermi level $\dote{F}$.

The second contribution derives from the thermal variations in the phonon distribution, providing accelerating forces involving the processes $\sum_{ss'}k_B\beta^2\omega_{\bfq s}(1+s\omega_{\perp\bfq}\cos\phi_q)[2\sinh(s'\beta\omega_{\bfq s}/2)(\dote{\bfq}+s'\omega_{\bfq s}-\bfk\cdot\bfq)]^{-2}(\bfr-\bfr_0)\cdot\nabla T$. The factor $s'$ makes reference to the absorption ($s'=-1$) and emission ($s'=1)$ processes, whereas $s$ refers to the effective chirality. Ultimately, the phonon driven contribution to the drift current results from the net of the two opposite chiralities, c.f. Eq. \eqref{eq-D}. The theoretical model predicts that the induced spin-currents should result in a spin accumulation at some interface surface intersecting the current, in agreement with recent experimental observations \cite{JunLiu}.

In conclusion, it has been demonstrated that the angular momentum carried by chiral phonons may be transferred to electrons via a phase exchange. The phonon chirality translates to the electron spin in the sense that it generates a non-trivial spin-texture including circulating spin-currents.

\acknowledgements
The author thanks the CISSors for helpful and constructive comments. The author, moreover, thanks R. Arouca, A. Bergman, L. Barreto Braz, O. Gr\aa n\"as, S. Kahlh\"ofer, L. Nordstr\"om, M. Shiranzaei, A. Sisman, and P. Thunstr\"om for discussions. Financial support from Vetenskapsr\aa det and Stiftelsen Olle Engkvist Byggm\"astare is acknowledged.

\bibliography{CPISPref}

\begin{thebibliography}{41}%
\makeatletter
\providecommand \@ifxundefined [1]{%
 \@ifx{#1\undefined}
}%
\providecommand \@ifnum [1]{%
 \ifnum #1\expandafter \@firstoftwo
 \else \expandafter \@secondoftwo
 \fi
}%
\providecommand \@ifx [1]{%
 \ifx #1\expandafter \@firstoftwo
 \else \expandafter \@secondoftwo
 \fi
}%
\providecommand \natexlab [1]{#1}%
\providecommand \enquote  [1]{``#1''}%
\providecommand \bibnamefont  [1]{#1}%
\providecommand \bibfnamefont [1]{#1}%
\providecommand \citenamefont [1]{#1}%
\providecommand \href@noop [0]{\@secondoftwo}%
\providecommand \href [0]{\begingroup \@sanitize@url \@href}%
\providecommand \@href[1]{\@@startlink{#1}\@@href}%
\providecommand \@@href[1]{\endgroup#1\@@endlink}%
\providecommand \@sanitize@url [0]{\catcode `\\12\catcode `\$12\catcode
  `\&12\catcode `\#12\catcode `\^12\catcode `\_12\catcode `\%12\relax}%
\providecommand \@@startlink[1]{}%
\providecommand \@@endlink[0]{}%
\providecommand \url  [0]{\begingroup\@sanitize@url \@url }%
\providecommand \@url [1]{\endgroup\@href {#1}{\urlprefix }}%
\providecommand \urlprefix  [0]{URL }%
\providecommand \Eprint [0]{\href }%
\providecommand \doibase [0]{https://doi.org/}%
\providecommand \selectlanguage [0]{\@gobble}%
\providecommand \bibinfo  [0]{\@secondoftwo}%
\providecommand \bibfield  [0]{\@secondoftwo}%
\providecommand \translation [1]{[#1]}%
\providecommand \BibitemOpen [0]{}%
\providecommand \bibitemStop [0]{}%
\providecommand \bibitemNoStop [0]{.\EOS\space}%
\providecommand \EOS [0]{\spacefactor3000\relax}%
\providecommand \BibitemShut  [1]{\csname bibitem#1\endcsname}%
\let\auto@bib@innerbib\@empty
%</preamble>
\bibitem [{\citenamefont {Strohm}\ \emph {et~al.}(2005)\citenamefont {Strohm},
  \citenamefont {Rikken},\ and\ \citenamefont {Wyder}}]{PhysRevLett.95.155901}%
  \BibitemOpen
  \bibfield  {author} {\bibinfo {author} {\bibfnamefont {C.}~\bibnamefont
  {Strohm}}, \bibinfo {author} {\bibfnamefont {G.~L. J.~A.}\ \bibnamefont
  {Rikken}},\ and\ \bibinfo {author} {\bibfnamefont {P.}~\bibnamefont
  {Wyder}},\ }\bibfield  {title} {\bibinfo {title} {Phenomenological evidence
  for the phonon hall effect},\ }\href
  {https://doi.org/10.1103/PhysRevLett.95.155901} {\bibfield  {journal}
  {\bibinfo  {journal} {Phys. Rev. Lett.}\ }\textbf {\bibinfo {volume} {95}},\
  \bibinfo {pages} {155901} (\bibinfo {year} {2005})}\BibitemShut {NoStop}%
\bibitem [{\citenamefont {Sheng}\ \emph {et~al.}(2006)\citenamefont {Sheng},
  \citenamefont {Sheng},\ and\ \citenamefont {Ting}}]{PhysRevLett.96.155901}%
  \BibitemOpen
  \bibfield  {author} {\bibinfo {author} {\bibfnamefont {L.}~\bibnamefont
  {Sheng}}, \bibinfo {author} {\bibfnamefont {D.~N.}\ \bibnamefont {Sheng}},\
  and\ \bibinfo {author} {\bibfnamefont {C.~S.}\ \bibnamefont {Ting}},\
  }\bibfield  {title} {\bibinfo {title} {Theory of the phonon hall effect in
  paramagnetic dielectrics},\ }\href
  {https://doi.org/10.1103/PhysRevLett.96.155901} {\bibfield  {journal}
  {\bibinfo  {journal} {Phys. Rev. Lett.}\ }\textbf {\bibinfo {volume} {96}},\
  \bibinfo {pages} {155901} (\bibinfo {year} {2006})}\BibitemShut {NoStop}%
\bibitem [{\citenamefont {Qin}\ \emph {et~al.}(2012)\citenamefont {Qin},
  \citenamefont {Zhou},\ and\ \citenamefont {Shi}}]{PhysRevB.86.104305}%
  \BibitemOpen
  \bibfield  {author} {\bibinfo {author} {\bibfnamefont {T.}~\bibnamefont
  {Qin}}, \bibinfo {author} {\bibfnamefont {J.}~\bibnamefont {Zhou}},\ and\
  \bibinfo {author} {\bibfnamefont {J.}~\bibnamefont {Shi}},\ }\bibfield
  {title} {\bibinfo {title} {Berry curvature and the phonon hall effect},\
  }\href {https://doi.org/10.1103/PhysRevB.86.104305} {\bibfield  {journal}
  {\bibinfo  {journal} {Phys. Rev. B}\ }\textbf {\bibinfo {volume} {86}},\
  \bibinfo {pages} {104305} (\bibinfo {year} {2012})}\BibitemShut {NoStop}%
\bibitem [{\citenamefont {Kagan}\ and\ \citenamefont
  {Maksimov}(2008)}]{PhysRevLett.100.145902}%
  \BibitemOpen
  \bibfield  {author} {\bibinfo {author} {\bibfnamefont {Y.}~\bibnamefont
  {Kagan}}\ and\ \bibinfo {author} {\bibfnamefont {L.~A.}\ \bibnamefont
  {Maksimov}},\ }\bibfield  {title} {\bibinfo {title} {Anomalous hall effect
  for the phonon heat conductivity in paramagnetic dielectrics},\ }\href
  {https://doi.org/10.1103/PhysRevLett.100.145902} {\bibfield  {journal}
  {\bibinfo  {journal} {Phys. Rev. Lett.}\ }\textbf {\bibinfo {volume} {100}},\
  \bibinfo {pages} {145902} (\bibinfo {year} {2008})}\BibitemShut {NoStop}%
\bibitem [{\citenamefont {Zhang}\ \emph {et~al.}(2010)\citenamefont {Zhang},
  \citenamefont {Ren}, \citenamefont {Wang},\ and\ \citenamefont
  {Li}}]{PhysRevLett.105.225901}%
  \BibitemOpen
  \bibfield  {author} {\bibinfo {author} {\bibfnamefont {L.}~\bibnamefont
  {Zhang}}, \bibinfo {author} {\bibfnamefont {J.}~\bibnamefont {Ren}}, \bibinfo
  {author} {\bibfnamefont {J.-S.}\ \bibnamefont {Wang}},\ and\ \bibinfo
  {author} {\bibfnamefont {B.}~\bibnamefont {Li}},\ }\bibfield  {title}
  {\bibinfo {title} {Topological nature of the phonon hall effect},\ }\href
  {https://doi.org/10.1103/PhysRevLett.105.225901} {\bibfield  {journal}
  {\bibinfo  {journal} {Phys. Rev. Lett.}\ }\textbf {\bibinfo {volume} {105}},\
  \bibinfo {pages} {225901} (\bibinfo {year} {2010})}\BibitemShut {NoStop}%
\bibitem [{\citenamefont {Mori}\ \emph {et~al.}(2014)\citenamefont {Mori},
  \citenamefont {Spencer-Smith}, \citenamefont {Sushkov},\ and\ \citenamefont
  {Maekawa}}]{PhysRevLett.113.265901}%
  \BibitemOpen
  \bibfield  {author} {\bibinfo {author} {\bibfnamefont {M.}~\bibnamefont
  {Mori}}, \bibinfo {author} {\bibfnamefont {A.}~\bibnamefont {Spencer-Smith}},
  \bibinfo {author} {\bibfnamefont {O.~P.}\ \bibnamefont {Sushkov}},\ and\
  \bibinfo {author} {\bibfnamefont {S.}~\bibnamefont {Maekawa}},\ }\bibfield
  {title} {\bibinfo {title} {Origin of the phonon hall effect in rare-earth
  garnets},\ }\href {https://doi.org/10.1103/PhysRevLett.113.265901} {\bibfield
   {journal} {\bibinfo  {journal} {Phys. Rev. Lett.}\ }\textbf {\bibinfo
  {volume} {113}},\ \bibinfo {pages} {265901} (\bibinfo {year}
  {2014})}\BibitemShut {NoStop}%
\bibitem [{\citenamefont {Park}\ and\ \citenamefont
  {Yang}(2020)}]{NanoLett.20.7694}%
  \BibitemOpen
  \bibfield  {author} {\bibinfo {author} {\bibfnamefont {S.}~\bibnamefont
  {Park}}\ and\ \bibinfo {author} {\bibfnamefont {B.-J.}\ \bibnamefont
  {Yang}},\ }\bibfield  {title} {\bibinfo {title} {Phonon angular momentum hall
  effect},\ }\href {https://doi.org/10.1021/acs.nanolett.0c03220} {\bibfield
  {journal} {\bibinfo  {journal} {Nano Lett.}\ }\textbf {\bibinfo {volume}
  {20}},\ \bibinfo {pages} {7694} (\bibinfo {year} {2020})}\BibitemShut
  {NoStop}%
\bibitem [{\citenamefont {Garanin}\ and\ \citenamefont
  {Chudnovsky}(2015)}]{PhysRevB.92.024421}%
  \BibitemOpen
  \bibfield  {author} {\bibinfo {author} {\bibfnamefont {D.~A.}\ \bibnamefont
  {Garanin}}\ and\ \bibinfo {author} {\bibfnamefont {E.~M.}\ \bibnamefont
  {Chudnovsky}},\ }\bibfield  {title} {\bibinfo {title} {Angular momentum in
  spin-phonon processes},\ }\href {https://doi.org/10.1103/PhysRevB.92.024421}
  {\bibfield  {journal} {\bibinfo  {journal} {Phys. Rev. B}\ }\textbf {\bibinfo
  {volume} {92}},\ \bibinfo {pages} {024421} (\bibinfo {year}
  {2015})}\BibitemShut {NoStop}%
\bibitem [{\citenamefont {Streib}\ \emph {et~al.}(2018)\citenamefont {Streib},
  \citenamefont {Keshtgar},\ and\ \citenamefont
  {Bauer}}]{PhysRevLett.121.027202}%
  \BibitemOpen
  \bibfield  {author} {\bibinfo {author} {\bibfnamefont {S.}~\bibnamefont
  {Streib}}, \bibinfo {author} {\bibfnamefont {H.}~\bibnamefont {Keshtgar}},\
  and\ \bibinfo {author} {\bibfnamefont {G.~E.~W.}\ \bibnamefont {Bauer}},\
  }\bibfield  {title} {\bibinfo {title} {Damping of magnetization dynamics by
  phonon pumping},\ }\href {https://doi.org/10.1103/PhysRevLett.121.027202}
  {\bibfield  {journal} {\bibinfo  {journal} {Phys. Rev. Lett.}\ }\textbf
  {\bibinfo {volume} {121}},\ \bibinfo {pages} {027202} (\bibinfo {year}
  {2018})}\BibitemShut {NoStop}%
\bibitem [{\citenamefont {Nakane}\ and\ \citenamefont
  {Kohno}(2018)}]{PhysRevB.97.174403}%
  \BibitemOpen
  \bibfield  {author} {\bibinfo {author} {\bibfnamefont {J.~J.}\ \bibnamefont
  {Nakane}}\ and\ \bibinfo {author} {\bibfnamefont {H.}~\bibnamefont {Kohno}},\
  }\bibfield  {title} {\bibinfo {title} {Angular momentum of phonons and its
  application to single-spin relaxation},\ }\href
  {https://doi.org/10.1103/PhysRevB.97.174403} {\bibfield  {journal} {\bibinfo
  {journal} {Phys. Rev. B}\ }\textbf {\bibinfo {volume} {97}},\ \bibinfo
  {pages} {174403} (\bibinfo {year} {2018})}\BibitemShut {NoStop}%
\bibitem [{\citenamefont {Bistoni}\ \emph {et~al.}(2021)\citenamefont
  {Bistoni}, \citenamefont {Mauri},\ and\ \citenamefont
  {Calandra}}]{PhysRevLett.126.225703}%
  \BibitemOpen
  \bibfield  {author} {\bibinfo {author} {\bibfnamefont {O.}~\bibnamefont
  {Bistoni}}, \bibinfo {author} {\bibfnamefont {F.}~\bibnamefont {Mauri}},\
  and\ \bibinfo {author} {\bibfnamefont {M.}~\bibnamefont {Calandra}},\
  }\bibfield  {title} {\bibinfo {title} {Intrinsic vibrational angular momentum
  from nonadiabatic effects in noncollinear magnetic molecules},\ }\href
  {https://doi.org/10.1103/PhysRevLett.126.225703} {\bibfield  {journal}
  {\bibinfo  {journal} {Phys. Rev. Lett.}\ }\textbf {\bibinfo {volume} {126}},\
  \bibinfo {pages} {225703} (\bibinfo {year} {2021})}\BibitemShut {NoStop}%
\bibitem [{\citenamefont {Zhang}\ and\ \citenamefont
  {Niu}(2014)}]{PhysRevLett.112.085503}%
  \BibitemOpen
  \bibfield  {author} {\bibinfo {author} {\bibfnamefont {L.}~\bibnamefont
  {Zhang}}\ and\ \bibinfo {author} {\bibfnamefont {Q.}~\bibnamefont {Niu}},\
  }\bibfield  {title} {\bibinfo {title} {Angular momentum of phonons and the
  einstein--de haas effect},\ }\href
  {https://doi.org/10.1103/PhysRevLett.112.085503} {\bibfield  {journal}
  {\bibinfo  {journal} {Phys. Rev. Lett.}\ }\textbf {\bibinfo {volume} {112}},\
  \bibinfo {pages} {085503} (\bibinfo {year} {2014})}\BibitemShut {NoStop}%
\bibitem [{\citenamefont {Mentink}\ \emph {et~al.}(2019)\citenamefont
  {Mentink}, \citenamefont {Katsnelson},\ and\ \citenamefont
  {Lemeshko}}]{PhysRevB.99.064428}%
  \BibitemOpen
  \bibfield  {author} {\bibinfo {author} {\bibfnamefont {J.~H.}\ \bibnamefont
  {Mentink}}, \bibinfo {author} {\bibfnamefont {M.~I.}\ \bibnamefont
  {Katsnelson}},\ and\ \bibinfo {author} {\bibfnamefont {M.}~\bibnamefont
  {Lemeshko}},\ }\bibfield  {title} {\bibinfo {title} {Quantum many-body
  dynamics of the einstein--de haas effect},\ }\href
  {https://doi.org/10.1103/PhysRevB.99.064428} {\bibfield  {journal} {\bibinfo
  {journal} {Phys. Rev. B}\ }\textbf {\bibinfo {volume} {99}},\ \bibinfo
  {pages} {064428} (\bibinfo {year} {2019})}\BibitemShut {NoStop}%
\bibitem [{\citenamefont {Dornes}\ \emph {et~al.}(2019)\citenamefont {Dornes},
  \citenamefont {Acremann}, \citenamefont {Savoini}, \citenamefont {Kubli},
  \citenamefont {Neugebauer}, \citenamefont {Abreu}, \citenamefont {Huber},
  \citenamefont {Lantz}, \citenamefont {Vaz}, \citenamefont {Lemke},
  \citenamefont {Bothschafter}, \citenamefont {Porer}, \citenamefont
  {Esposito}, \citenamefont {Rettig}, \citenamefont {Buzzi}, \citenamefont
  {Alberca}, \citenamefont {Windsor}, \citenamefont {Beaud}, \citenamefont
  {Staub}, \citenamefont {Zhu}, \citenamefont {Song}, \citenamefont {Glownia},\
  and\ \citenamefont {Johnson}}]{Nature.565.209}%
  \BibitemOpen
  \bibfield  {author} {\bibinfo {author} {\bibfnamefont {C.}~\bibnamefont
  {Dornes}}, \bibinfo {author} {\bibfnamefont {Y.}~\bibnamefont {Acremann}},
  \bibinfo {author} {\bibfnamefont {M.}~\bibnamefont {Savoini}}, \bibinfo
  {author} {\bibfnamefont {M.}~\bibnamefont {Kubli}}, \bibinfo {author}
  {\bibfnamefont {M.~J.}\ \bibnamefont {Neugebauer}}, \bibinfo {author}
  {\bibfnamefont {E.}~\bibnamefont {Abreu}}, \bibinfo {author} {\bibfnamefont
  {L.}~\bibnamefont {Huber}}, \bibinfo {author} {\bibfnamefont
  {G.}~\bibnamefont {Lantz}}, \bibinfo {author} {\bibfnamefont {C.~A.~F.}\
  \bibnamefont {Vaz}}, \bibinfo {author} {\bibfnamefont {H.}~\bibnamefont
  {Lemke}}, \bibinfo {author} {\bibfnamefont {E.~M.}\ \bibnamefont
  {Bothschafter}}, \bibinfo {author} {\bibfnamefont {M.}~\bibnamefont {Porer}},
  \bibinfo {author} {\bibfnamefont {V.}~\bibnamefont {Esposito}}, \bibinfo
  {author} {\bibfnamefont {L.}~\bibnamefont {Rettig}}, \bibinfo {author}
  {\bibfnamefont {M.}~\bibnamefont {Buzzi}}, \bibinfo {author} {\bibfnamefont
  {A.}~\bibnamefont {Alberca}}, \bibinfo {author} {\bibfnamefont {Y.~W.}\
  \bibnamefont {Windsor}}, \bibinfo {author} {\bibfnamefont {P.}~\bibnamefont
  {Beaud}}, \bibinfo {author} {\bibfnamefont {U.}~\bibnamefont {Staub}},
  \bibinfo {author} {\bibfnamefont {D.}~\bibnamefont {Zhu}}, \bibinfo {author}
  {\bibfnamefont {S.}~\bibnamefont {Song}}, \bibinfo {author} {\bibfnamefont
  {J.~M.}\ \bibnamefont {Glownia}},\ and\ \bibinfo {author} {\bibfnamefont
  {S.~L.}\ \bibnamefont {Johnson}},\ }\bibfield  {title} {\bibinfo {title} {The
  ultrafast einstein--de haas effect},\ }\href
  {https://doi.org/10.1038/s41586-018-0822-7} {\bibfield  {journal} {\bibinfo
  {journal} {Nature}\ }\textbf {\bibinfo {volume} {565}},\ \bibinfo {pages}
  {209} (\bibinfo {year} {2019})}\BibitemShut {NoStop}%
\bibitem [{\citenamefont {Fransson}\ \emph {et~al.}(2017)\citenamefont
  {Fransson}, \citenamefont {Thonig}, \citenamefont {Bessarab}, \citenamefont
  {Bhattacharjee}, \citenamefont {Hellsvik},\ and\ \citenamefont
  {Nordstr\"om}}]{PhysRevMaterials.1.074404}%
  \BibitemOpen
  \bibfield  {author} {\bibinfo {author} {\bibfnamefont {J.}~\bibnamefont
  {Fransson}}, \bibinfo {author} {\bibfnamefont {D.}~\bibnamefont {Thonig}},
  \bibinfo {author} {\bibfnamefont {P.~F.}\ \bibnamefont {Bessarab}}, \bibinfo
  {author} {\bibfnamefont {S.}~\bibnamefont {Bhattacharjee}}, \bibinfo {author}
  {\bibfnamefont {J.}~\bibnamefont {Hellsvik}},\ and\ \bibinfo {author}
  {\bibfnamefont {L.}~\bibnamefont {Nordstr\"om}},\ }\bibfield  {title}
  {\bibinfo {title} {Microscopic theory for coupled atomistic magnetization and
  lattice dynamics},\ }\href
  {https://doi.org/10.1103/PhysRevMaterials.1.074404} {\bibfield  {journal}
  {\bibinfo  {journal} {Phys. Rev. Materials}\ }\textbf {\bibinfo {volume}
  {1}},\ \bibinfo {pages} {074404} (\bibinfo {year} {2017})}\BibitemShut
  {NoStop}%
\bibitem [{\citenamefont {Bennett}\ \emph {et~al.}(2013)\citenamefont
  {Bennett}, \citenamefont {Yao}, \citenamefont {Otterbach}, \citenamefont
  {Zoller}, \citenamefont {Rabl},\ and\ \citenamefont
  {Lukin}}]{PhysRevLett.110.156402}%
  \BibitemOpen
  \bibfield  {author} {\bibinfo {author} {\bibfnamefont {S.~D.}\ \bibnamefont
  {Bennett}}, \bibinfo {author} {\bibfnamefont {N.~Y.}\ \bibnamefont {Yao}},
  \bibinfo {author} {\bibfnamefont {J.}~\bibnamefont {Otterbach}}, \bibinfo
  {author} {\bibfnamefont {P.}~\bibnamefont {Zoller}}, \bibinfo {author}
  {\bibfnamefont {P.}~\bibnamefont {Rabl}},\ and\ \bibinfo {author}
  {\bibfnamefont {M.~D.}\ \bibnamefont {Lukin}},\ }\bibfield  {title} {\bibinfo
  {title} {Phonon-induced spin-spin interactions in diamond nanostructures:
  Application to spin squeezing},\ }\href
  {https://doi.org/10.1103/PhysRevLett.110.156402} {\bibfield  {journal}
  {\bibinfo  {journal} {Phys. Rev. Lett.}\ }\textbf {\bibinfo {volume} {110}},\
  \bibinfo {pages} {156402} (\bibinfo {year} {2013})}\BibitemShut {NoStop}%
\bibitem [{\citenamefont {Hamada}\ \emph {et~al.}(2018)\citenamefont {Hamada},
  \citenamefont {Minamitani}, \citenamefont {Hirayama},\ and\ \citenamefont
  {Murakami}}]{PhysRevLett.121.175301}%
  \BibitemOpen
  \bibfield  {author} {\bibinfo {author} {\bibfnamefont {M.}~\bibnamefont
  {Hamada}}, \bibinfo {author} {\bibfnamefont {E.}~\bibnamefont {Minamitani}},
  \bibinfo {author} {\bibfnamefont {M.}~\bibnamefont {Hirayama}},\ and\
  \bibinfo {author} {\bibfnamefont {S.}~\bibnamefont {Murakami}},\ }\bibfield
  {title} {\bibinfo {title} {Phonon angular momentum induced by the temperature
  gradient},\ }\href {https://doi.org/10.1103/PhysRevLett.121.175301}
  {\bibfield  {journal} {\bibinfo  {journal} {Phys. Rev. Lett.}\ }\textbf
  {\bibinfo {volume} {121}},\ \bibinfo {pages} {175301} (\bibinfo {year}
  {2018})}\BibitemShut {NoStop}%
\bibitem [{\citenamefont {Juraschek}\ and\ \citenamefont
  {Spaldin}(2019)}]{PhysRevMaterials.3.064405}%
  \BibitemOpen
  \bibfield  {author} {\bibinfo {author} {\bibfnamefont {D.~M.}\ \bibnamefont
  {Juraschek}}\ and\ \bibinfo {author} {\bibfnamefont {N.~A.}\ \bibnamefont
  {Spaldin}},\ }\bibfield  {title} {\bibinfo {title} {Orbital magnetic moments
  of phonons},\ }\href {https://doi.org/10.1103/PhysRevMaterials.3.064405}
  {\bibfield  {journal} {\bibinfo  {journal} {Phys. Rev. Materials}\ }\textbf
  {\bibinfo {volume} {3}},\ \bibinfo {pages} {064405} (\bibinfo {year}
  {2019})}\BibitemShut {NoStop}%
\bibitem [{\citenamefont {Juraschek}\ \emph {et~al.}(2020)\citenamefont
  {Juraschek}, \citenamefont {Narang},\ and\ \citenamefont
  {Spaldin}}]{PhysRevResearch.2.043035}%
  \BibitemOpen
  \bibfield  {author} {\bibinfo {author} {\bibfnamefont {D.~M.}\ \bibnamefont
  {Juraschek}}, \bibinfo {author} {\bibfnamefont {P.}~\bibnamefont {Narang}},\
  and\ \bibinfo {author} {\bibfnamefont {N.~A.}\ \bibnamefont {Spaldin}},\
  }\bibfield  {title} {\bibinfo {title} {Phono-magnetic analogs to
  opto-magnetic effects},\ }\href
  {https://doi.org/10.1103/PhysRevResearch.2.043035} {\bibfield  {journal}
  {\bibinfo  {journal} {Phys. Rev. Research}\ }\textbf {\bibinfo {volume}
  {2}},\ \bibinfo {pages} {043035} (\bibinfo {year} {2020})}\BibitemShut
  {NoStop}%
\bibitem [{\citenamefont {Juraschek}\ \emph {et~al.}(2022)\citenamefont
  {Juraschek}, \citenamefont {Neuman},\ and\ \citenamefont
  {Narang}}]{PhysRevResearch.4.013129}%
  \BibitemOpen
  \bibfield  {author} {\bibinfo {author} {\bibfnamefont {D.~M.}\ \bibnamefont
  {Juraschek}}, \bibinfo {author} {\bibfnamefont {T.~c.~v.}\ \bibnamefont
  {Neuman}},\ and\ \bibinfo {author} {\bibfnamefont {P.}~\bibnamefont
  {Narang}},\ }\bibfield  {title} {\bibinfo {title} {Giant effective magnetic
  fields from optically driven chiral phonons in $4f$ paramagnets},\ }\href
  {https://doi.org/10.1103/PhysRevResearch.4.013129} {\bibfield  {journal}
  {\bibinfo  {journal} {Phys. Rev. Research}\ }\textbf {\bibinfo {volume}
  {4}},\ \bibinfo {pages} {013129} (\bibinfo {year} {2022})}\BibitemShut
  {NoStop}%
\bibitem [{\citenamefont {Zhu}\ \emph {et~al.}(2018)\citenamefont {Zhu},
  \citenamefont {Yi}, \citenamefont {Li}, \citenamefont {Xiao}, \citenamefont
  {Zhang}, \citenamefont {Yang}, \citenamefont {Kaindl}, \citenamefont {Li},
  \citenamefont {Wang},\ and\ \citenamefont {Zhang}}]{Science.359.579}%
  \BibitemOpen
  \bibfield  {author} {\bibinfo {author} {\bibfnamefont {H.}~\bibnamefont
  {Zhu}}, \bibinfo {author} {\bibfnamefont {J.}~\bibnamefont {Yi}}, \bibinfo
  {author} {\bibfnamefont {M.-Y.}\ \bibnamefont {Li}}, \bibinfo {author}
  {\bibfnamefont {J.}~\bibnamefont {Xiao}}, \bibinfo {author} {\bibfnamefont
  {L.}~\bibnamefont {Zhang}}, \bibinfo {author} {\bibfnamefont {C.-W.}\
  \bibnamefont {Yang}}, \bibinfo {author} {\bibfnamefont {R.~A.}\ \bibnamefont
  {Kaindl}}, \bibinfo {author} {\bibfnamefont {L.-J.}\ \bibnamefont {Li}},
  \bibinfo {author} {\bibfnamefont {Y.}~\bibnamefont {Wang}},\ and\ \bibinfo
  {author} {\bibfnamefont {X.}~\bibnamefont {Zhang}},\ }\bibfield  {title}
  {\bibinfo {title} {Observation of chiral phonons},\ }\href
  {https://doi.org/10.1126/science.aar2711} {\bibfield  {journal} {\bibinfo
  {journal} {Science}\ }\textbf {\bibinfo {volume} {359}},\ \bibinfo {pages}
  {579} (\bibinfo {year} {2018})}\BibitemShut {NoStop}%
\bibitem [{\citenamefont {Yin}\ \emph {et~al.}(2021)\citenamefont {Yin},
  \citenamefont {Ulman}, \citenamefont {Liu}, \citenamefont {Granados~del
  {\'A}guila}, \citenamefont {Huang}, \citenamefont {Zhang}, \citenamefont
  {Serra}, \citenamefont {Sedmidubsky}, \citenamefont {Sofer}, \citenamefont
  {Quek},\ and\ \citenamefont {Xiong}}]{AdvMater.33.2101618}%
  \BibitemOpen
  \bibfield  {author} {\bibinfo {author} {\bibfnamefont {T.}~\bibnamefont
  {Yin}}, \bibinfo {author} {\bibfnamefont {K.~A.}\ \bibnamefont {Ulman}},
  \bibinfo {author} {\bibfnamefont {S.}~\bibnamefont {Liu}}, \bibinfo {author}
  {\bibfnamefont {A.}~\bibnamefont {Granados~del {\'A}guila}}, \bibinfo
  {author} {\bibfnamefont {Y.}~\bibnamefont {Huang}}, \bibinfo {author}
  {\bibfnamefont {L.}~\bibnamefont {Zhang}}, \bibinfo {author} {\bibfnamefont
  {M.}~\bibnamefont {Serra}}, \bibinfo {author} {\bibfnamefont
  {D.}~\bibnamefont {Sedmidubsky}}, \bibinfo {author} {\bibfnamefont
  {Z.}~\bibnamefont {Sofer}}, \bibinfo {author} {\bibfnamefont {S.~Y.}\
  \bibnamefont {Quek}},\ and\ \bibinfo {author} {\bibfnamefont
  {Q.}~\bibnamefont {Xiong}},\ }\bibfield  {title} {\bibinfo {title} {Chiral
  phonons and giant magneto-optical effect in crbr3 2d magnet},\ }\href
  {https://doi.org/https://doi.org/10.1002/adma.202101618} {\bibfield
  {journal} {\bibinfo  {journal} {Adv. Mater.}\ }\textbf {\bibinfo {volume}
  {33}},\ \bibinfo {pages} {2101618} (\bibinfo {year} {2021})}\BibitemShut
  {NoStop}%
\bibitem [{\citenamefont {Jeong}\ \emph {et~al.}(2022)\citenamefont {Jeong},
  \citenamefont {Kim}, \citenamefont {Seo}, \citenamefont {Park}, \citenamefont
  {Jeong}, \citenamefont {Kim}, \citenamefont {Lauter}, \citenamefont {Egami},
  \citenamefont {Han},\ and\ \citenamefont {Choi}}]{SciAdv.8.eabm4005}%
  \BibitemOpen
  \bibfield  {author} {\bibinfo {author} {\bibfnamefont {S.~G.}\ \bibnamefont
  {Jeong}}, \bibinfo {author} {\bibfnamefont {J.}~\bibnamefont {Kim}}, \bibinfo
  {author} {\bibfnamefont {A.}~\bibnamefont {Seo}}, \bibinfo {author}
  {\bibfnamefont {S.}~\bibnamefont {Park}}, \bibinfo {author} {\bibfnamefont
  {H.~Y.}\ \bibnamefont {Jeong}}, \bibinfo {author} {\bibfnamefont {Y.-M.}\
  \bibnamefont {Kim}}, \bibinfo {author} {\bibfnamefont {V.}~\bibnamefont
  {Lauter}}, \bibinfo {author} {\bibfnamefont {T.}~\bibnamefont {Egami}},
  \bibinfo {author} {\bibfnamefont {J.~H.}\ \bibnamefont {Han}},\ and\ \bibinfo
  {author} {\bibfnamefont {W.~S.}\ \bibnamefont {Choi}},\ }\bibfield  {title}
  {\bibinfo {title} {Unconventional interlayer exchange coupling via chiral
  phonons in synthetic magnetic oxide heterostructures},\ }\href
  {https://doi.org/10.1126/sciadv.abm4005} {\bibfield  {journal} {\bibinfo
  {journal} {Sci. Adv.}\ }\textbf {\bibinfo {volume} {8}},\ \bibinfo {pages}
  {eabm4005} (\bibinfo {year} {2022})}\BibitemShut {NoStop}%
\bibitem [{\citenamefont {Chen}\ \emph {et~al.}(2019)\citenamefont {Chen},
  \citenamefont {Lu}, \citenamefont {Dubey}, \citenamefont {Yao}, \citenamefont
  {Liu}, \citenamefont {Wang}, \citenamefont {Xiong}, \citenamefont {Zhang},\
  and\ \citenamefont {Srivastava}}]{NaturePhys.15.221}%
  \BibitemOpen
  \bibfield  {author} {\bibinfo {author} {\bibfnamefont {X.}~\bibnamefont
  {Chen}}, \bibinfo {author} {\bibfnamefont {X.}~\bibnamefont {Lu}}, \bibinfo
  {author} {\bibfnamefont {S.}~\bibnamefont {Dubey}}, \bibinfo {author}
  {\bibfnamefont {Q.}~\bibnamefont {Yao}}, \bibinfo {author} {\bibfnamefont
  {S.}~\bibnamefont {Liu}}, \bibinfo {author} {\bibfnamefont {X.}~\bibnamefont
  {Wang}}, \bibinfo {author} {\bibfnamefont {Q.}~\bibnamefont {Xiong}},
  \bibinfo {author} {\bibfnamefont {L.}~\bibnamefont {Zhang}},\ and\ \bibinfo
  {author} {\bibfnamefont {A.}~\bibnamefont {Srivastava}},\ }\bibfield  {title}
  {\bibinfo {title} {Entanglement of single-photons and chiral phonons in
  atomically thin wse2},\ }\href {https://doi.org/10.1038/s41567-018-0366-7}
  {\bibfield  {journal} {\bibinfo  {journal} {Nat. Phys.}\ }\textbf {\bibinfo
  {volume} {15}},\ \bibinfo {pages} {221} (\bibinfo {year} {2019})}\BibitemShut
  {NoStop}%
\bibitem [{\citenamefont {Baydin}\ \emph {et~al.}(2022)\citenamefont {Baydin},
  \citenamefont {Hernandez}, \citenamefont {Rodriguez-Vega}, \citenamefont
  {Okazaki}, \citenamefont {Tay}, \citenamefont {Noe}, \citenamefont
  {Katayama}, \citenamefont {Takeda}, \citenamefont {Nojiri}, \citenamefont
  {Rappl}, \citenamefont {Abramof}, \citenamefont {Fiete},\ and\ \citenamefont
  {Kono}}]{PhysRevLett.128.075901}%
  \BibitemOpen
  \bibfield  {author} {\bibinfo {author} {\bibfnamefont {A.}~\bibnamefont
  {Baydin}}, \bibinfo {author} {\bibfnamefont {F.~G.~G.}\ \bibnamefont
  {Hernandez}}, \bibinfo {author} {\bibfnamefont {M.}~\bibnamefont
  {Rodriguez-Vega}}, \bibinfo {author} {\bibfnamefont {A.~K.}\ \bibnamefont
  {Okazaki}}, \bibinfo {author} {\bibfnamefont {F.}~\bibnamefont {Tay}},
  \bibinfo {author} {\bibfnamefont {G.~T.}\ \bibnamefont {Noe}}, \bibinfo
  {author} {\bibfnamefont {I.}~\bibnamefont {Katayama}}, \bibinfo {author}
  {\bibfnamefont {J.}~\bibnamefont {Takeda}}, \bibinfo {author} {\bibfnamefont
  {H.}~\bibnamefont {Nojiri}}, \bibinfo {author} {\bibfnamefont {P.~H.~O.}\
  \bibnamefont {Rappl}}, \bibinfo {author} {\bibfnamefont {E.}~\bibnamefont
  {Abramof}}, \bibinfo {author} {\bibfnamefont {G.~A.}\ \bibnamefont {Fiete}},\
  and\ \bibinfo {author} {\bibfnamefont {J.}~\bibnamefont {Kono}},\ }\bibfield
  {title} {\bibinfo {title} {Magnetic control of soft chiral phonons in pbte},\
  }\href {https://doi.org/10.1103/PhysRevLett.128.075901} {\bibfield  {journal}
  {\bibinfo  {journal} {Phys. Rev. Lett.}\ }\textbf {\bibinfo {volume} {128}},\
  \bibinfo {pages} {075901} (\bibinfo {year} {2022})}\BibitemShut {NoStop}%
\bibitem [{\citenamefont {Jaworski}\ \emph {et~al.}(2011)\citenamefont
  {Jaworski}, \citenamefont {Yang}, \citenamefont {Mack}, \citenamefont
  {Awschalom}, \citenamefont {Myers},\ and\ \citenamefont
  {Heremans}}]{PhysRevLett.106.186601}%
  \BibitemOpen
  \bibfield  {author} {\bibinfo {author} {\bibfnamefont {C.~M.}\ \bibnamefont
  {Jaworski}}, \bibinfo {author} {\bibfnamefont {J.}~\bibnamefont {Yang}},
  \bibinfo {author} {\bibfnamefont {S.}~\bibnamefont {Mack}}, \bibinfo {author}
  {\bibfnamefont {D.~D.}\ \bibnamefont {Awschalom}}, \bibinfo {author}
  {\bibfnamefont {R.~C.}\ \bibnamefont {Myers}},\ and\ \bibinfo {author}
  {\bibfnamefont {J.~P.}\ \bibnamefont {Heremans}},\ }\bibfield  {title}
  {\bibinfo {title} {Spin-seebeck effect: A phonon driven spin distribution},\
  }\href {https://doi.org/10.1103/PhysRevLett.106.186601} {\bibfield  {journal}
  {\bibinfo  {journal} {Phys. Rev. Lett.}\ }\textbf {\bibinfo {volume} {106}},\
  \bibinfo {pages} {186601} (\bibinfo {year} {2011})}\BibitemShut {NoStop}%
\bibitem [{\citenamefont {Lou}\ \emph {et~al.}(2018)\citenamefont {Lou},
  \citenamefont {{de Sousa Oliveira}}, \citenamefont {Tang}, \citenamefont
  {Greaney},\ and\ \citenamefont {Kumar}}]{SolidStateComm.283.37}%
  \BibitemOpen
  \bibfield  {author} {\bibinfo {author} {\bibfnamefont {P.~C.}\ \bibnamefont
  {Lou}}, \bibinfo {author} {\bibfnamefont {L.}~\bibnamefont {{de Sousa
  Oliveira}}}, \bibinfo {author} {\bibfnamefont {C.}~\bibnamefont {Tang}},
  \bibinfo {author} {\bibfnamefont {A.}~\bibnamefont {Greaney}},\ and\ \bibinfo
  {author} {\bibfnamefont {S.}~\bibnamefont {Kumar}},\ }\bibfield  {title}
  {\bibinfo {title} {Spin phonon interactions and magneto-thermal transport
  behavior in p-si},\ }\href
  {https://doi.org/https://doi.org/10.1016/j.ssc.2018.08.008} {\bibfield
  {journal} {\bibinfo  {journal} {Solid State Communications}\ }\textbf
  {\bibinfo {volume} {283}},\ \bibinfo {pages} {37} (\bibinfo {year}
  {2018})}\BibitemShut {NoStop}%
\bibitem [{\citenamefont {Du}\ \emph {et~al.}(2019)\citenamefont {Du},
  \citenamefont {Tang}, \citenamefont {Zhao}, \citenamefont {Li}, \citenamefont
  {Yang}, \citenamefont {Hu}, \citenamefont {Bai}, \citenamefont {Wang},
  \citenamefont {Watanabe}, \citenamefont {Taniguchi}, \citenamefont {Shi},
  \citenamefont {Yu}, \citenamefont {Bai}, \citenamefont {Hasan}, \citenamefont
  {Zhang},\ and\ \citenamefont {Sun}}]{AdvFunctMater.29.1904734}%
  \BibitemOpen
  \bibfield  {author} {\bibinfo {author} {\bibfnamefont {L.}~\bibnamefont
  {Du}}, \bibinfo {author} {\bibfnamefont {J.}~\bibnamefont {Tang}}, \bibinfo
  {author} {\bibfnamefont {Y.}~\bibnamefont {Zhao}}, \bibinfo {author}
  {\bibfnamefont {X.}~\bibnamefont {Li}}, \bibinfo {author} {\bibfnamefont
  {R.}~\bibnamefont {Yang}}, \bibinfo {author} {\bibfnamefont {X.}~\bibnamefont
  {Hu}}, \bibinfo {author} {\bibfnamefont {X.}~\bibnamefont {Bai}}, \bibinfo
  {author} {\bibfnamefont {X.}~\bibnamefont {Wang}}, \bibinfo {author}
  {\bibfnamefont {K.}~\bibnamefont {Watanabe}}, \bibinfo {author}
  {\bibfnamefont {T.}~\bibnamefont {Taniguchi}}, \bibinfo {author}
  {\bibfnamefont {D.}~\bibnamefont {Shi}}, \bibinfo {author} {\bibfnamefont
  {G.}~\bibnamefont {Yu}}, \bibinfo {author} {\bibfnamefont {X.}~\bibnamefont
  {Bai}}, \bibinfo {author} {\bibfnamefont {T.}~\bibnamefont {Hasan}}, \bibinfo
  {author} {\bibfnamefont {G.}~\bibnamefont {Zhang}},\ and\ \bibinfo {author}
  {\bibfnamefont {Z.}~\bibnamefont {Sun}},\ }\bibfield  {title} {\bibinfo
  {title} {Lattice dynamics, phonon chirality, and spin--phonon coupling in 2d
  itinerant ferromagnet fe3gete2},\ }\href
  {https://doi.org/https://doi.org/10.1002/adfm.201904734} {\bibfield
  {journal} {\bibinfo  {journal} {Adv. Funct. Mater.}\ }\textbf {\bibinfo
  {volume} {29}},\ \bibinfo {pages} {1904734} (\bibinfo {year}
  {2019})}\BibitemShut {NoStop}%
\bibitem [{\citenamefont {Overhauser}(1953)}]{PhysRev.89.689}%
  \BibitemOpen
  \bibfield  {author} {\bibinfo {author} {\bibfnamefont {A.~W.}\ \bibnamefont
  {Overhauser}},\ }\bibfield  {title} {\bibinfo {title} {Paramagnetic
  relaxation in metals},\ }\href {https://doi.org/10.1103/PhysRev.89.689}
  {\bibfield  {journal} {\bibinfo  {journal} {Phys. Rev.}\ }\textbf {\bibinfo
  {volume} {89}},\ \bibinfo {pages} {689} (\bibinfo {year} {1953})}\BibitemShut
  {NoStop}%
\bibitem [{\citenamefont {Elliott}(1954)}]{PhysRev.96.266}%
  \BibitemOpen
  \bibfield  {author} {\bibinfo {author} {\bibfnamefont {R.~J.}\ \bibnamefont
  {Elliott}},\ }\bibfield  {title} {\bibinfo {title} {Theory of the effect of
  spin-orbit coupling on magnetic resonance in some semiconductors},\ }\href
  {https://doi.org/10.1103/PhysRev.96.266} {\bibfield  {journal} {\bibinfo
  {journal} {Phys. Rev.}\ }\textbf {\bibinfo {volume} {96}},\ \bibinfo {pages}
  {266} (\bibinfo {year} {1954})}\BibitemShut {NoStop}%
\bibitem [{\citenamefont {Andreev}\ and\ \citenamefont
  {Gerasimenko}(1959)}]{JETP.35.846}%
  \BibitemOpen
  \bibfield  {author} {\bibinfo {author} {\bibfnamefont {V.~V.}\ \bibnamefont
  {Andreev}}\ and\ \bibinfo {author} {\bibfnamefont {V.~I.}\ \bibnamefont
  {Gerasimenko}},\ }\bibfield  {title} {\bibinfo {title} {On the theory of
  paramagnetic resonance and paramagnetic relaxation in metals},\ }\href
  {http://jetp.ras.ru/cgi-bin/dn/e_008_05_0846.pdf} {\bibfield  {journal}
  {\bibinfo  {journal} {Sov. Phys. JETP}\ }\textbf {\bibinfo {volume} {35}},\
  \bibinfo {pages} {846} (\bibinfo {year} {1959})}\BibitemShut {NoStop}%
\bibitem [{\citenamefont {Yafet}(1663)}]{Yafet1963}%
  \BibitemOpen
  \bibfield  {author} {\bibinfo {author} {\bibfnamefont {Y.}~\bibnamefont
  {Yafet}},\ }\href@noop {} {\emph {\bibinfo {title} {Solid State Phys.}}},\
  edited by\ \bibinfo {editor} {\bibfnamefont {F.}~\bibnamefont {Seitz}}\ and\
  \bibinfo {editor} {\bibfnamefont {D.}~\bibnamefont {Turnball}}\ (\bibinfo
  {publisher} {Academic},\ \bibinfo {year} {1663})\ pp.\ \bibinfo {pages}
  {1--98}\BibitemShut {NoStop}%
\bibitem [{\citenamefont {Baral}\ \emph {et~al.}(2016)\citenamefont {Baral},
  \citenamefont {Vollmar}, \citenamefont {Kaltenborn},\ and\ \citenamefont
  {Schneider}}]{NJP.18.023012}%
  \BibitemOpen
  \bibfield  {author} {\bibinfo {author} {\bibfnamefont {A.}~\bibnamefont
  {Baral}}, \bibinfo {author} {\bibfnamefont {S.}~\bibnamefont {Vollmar}},
  \bibinfo {author} {\bibfnamefont {S.}~\bibnamefont {Kaltenborn}},\ and\
  \bibinfo {author} {\bibfnamefont {H.~C.}\ \bibnamefont {Schneider}},\
  }\bibfield  {title} {\bibinfo {title} {Re-examination of the
  elliott{\textendash}yafet spin-relaxation mechanism},\ }\href
  {https://doi.org/10.1088/1367-2630/18/2/023012} {\bibfield  {journal}
  {\bibinfo  {journal} {New J. Phys.}\ }\textbf {\bibinfo {volume} {18}},\
  \bibinfo {pages} {023012} (\bibinfo {year} {2016})}\BibitemShut {NoStop}%
\bibitem [{\citenamefont {Fransson}(2020)}]{PhysRevB.102.235416}%
  \BibitemOpen
  \bibfield  {author} {\bibinfo {author} {\bibfnamefont {J.}~\bibnamefont
  {Fransson}},\ }\bibfield  {title} {\bibinfo {title} {Vibrational origin of
  exchange splitting and chiral-induced spin selectivity},\ }\href
  {https://doi.org/10.1103/PhysRevB.102.235416} {\bibfield  {journal} {\bibinfo
   {journal} {Phys. Rev. B}\ }\textbf {\bibinfo {volume} {102}},\ \bibinfo
  {pages} {235416} (\bibinfo {year} {2020})}\BibitemShut {NoStop}%
\bibitem [{\citenamefont {Du}\ \emph {et~al.}(2020)\citenamefont {Du},
  \citenamefont {Fu},\ and\ \citenamefont {Wu}}]{PhysRevB.102.035431}%
  \BibitemOpen
  \bibfield  {author} {\bibinfo {author} {\bibfnamefont {G.-F.}\ \bibnamefont
  {Du}}, \bibinfo {author} {\bibfnamefont {H.-H.}\ \bibnamefont {Fu}},\ and\
  \bibinfo {author} {\bibfnamefont {R.}~\bibnamefont {Wu}},\ }\bibfield
  {title} {\bibinfo {title} {Vibration-enhanced spin-selective transport of
  electrons in the dna double helix},\ }\href
  {https://doi.org/10.1103/PhysRevB.102.035431} {\bibfield  {journal} {\bibinfo
   {journal} {Phys. Rev. B}\ }\textbf {\bibinfo {volume} {102}},\ \bibinfo
  {pages} {035431} (\bibinfo {year} {2020})}\BibitemShut {NoStop}%
\bibitem [{\citenamefont {Zhang}\ \emph {et~al.}(2020)\citenamefont {Zhang},
  \citenamefont {Hao}, \citenamefont {Qin}, \citenamefont {Xie},\ and\
  \citenamefont {Qu}}]{PhysRevB.102.214303}%
  \BibitemOpen
  \bibfield  {author} {\bibinfo {author} {\bibfnamefont {L.}~\bibnamefont
  {Zhang}}, \bibinfo {author} {\bibfnamefont {Y.}~\bibnamefont {Hao}}, \bibinfo
  {author} {\bibfnamefont {W.}~\bibnamefont {Qin}}, \bibinfo {author}
  {\bibfnamefont {S.}~\bibnamefont {Xie}},\ and\ \bibinfo {author}
  {\bibfnamefont {F.}~\bibnamefont {Qu}},\ }\bibfield  {title} {\bibinfo
  {title} {Chiral-induced spin selectivity: A polaron transport model},\ }\href
  {https://doi.org/10.1103/PhysRevB.102.214303} {\bibfield  {journal} {\bibinfo
   {journal} {Phys. Rev. B}\ }\textbf {\bibinfo {volume} {102}},\ \bibinfo
  {pages} {214303} (\bibinfo {year} {2020})}\BibitemShut {NoStop}%
\bibitem [{\citenamefont {Fransson}(2021)}]{NanoLett.21.3026}%
  \BibitemOpen
  \bibfield  {author} {\bibinfo {author} {\bibfnamefont {J.}~\bibnamefont
  {Fransson}},\ }\bibfield  {title} {\bibinfo {title} {Charge redistribution
  and spin polarization driven by correlation induced electron exchange in
  chiral molecules},\ }\href {https://doi.org/10.1021/acs.nanolett.1c00183}
  {\bibfield  {journal} {\bibinfo  {journal} {Nano Lett.}\ }\textbf {\bibinfo
  {volume} {21}},\ \bibinfo {pages} {3026} (\bibinfo {year}
  {2021})}\BibitemShut {NoStop}%
\bibitem [{\citenamefont {Li}\ \emph {et~al.}(2021)\citenamefont {Li},
  \citenamefont {Zhong}, \citenamefont {Cheng}, \citenamefont {Chen},
  \citenamefont {Zhang},\ and\ \citenamefont {Zhou}}]{arXiv.2105.08485}%
  \BibitemOpen
  \bibfield  {author} {\bibinfo {author} {\bibfnamefont {X.}~\bibnamefont
  {Li}}, \bibinfo {author} {\bibfnamefont {J.}~\bibnamefont {Zhong}}, \bibinfo
  {author} {\bibfnamefont {J.}~\bibnamefont {Cheng}}, \bibinfo {author}
  {\bibfnamefont {H.}~\bibnamefont {Chen}}, \bibinfo {author} {\bibfnamefont
  {L.}~\bibnamefont {Zhang}},\ and\ \bibinfo {author} {\bibfnamefont
  {J.}~\bibnamefont {Zhou}},\ }\bibfield  {title} {\bibinfo {title} {Chiral
  phonon activated spin seebeck effect}\ }\href
  {https://doi.org/10.48550/ARXIV.2105.08485} {10.48550/ARXIV.2105.08485}
  (\bibinfo {year} {2021})\BibitemShut {NoStop}%
\bibitem [{\citenamefont {Kim}\ \emph {et~al.}(2022)\citenamefont {Kim},
  \citenamefont {Vetter}, \citenamefont {Yan}, \citenamefont {Yang},
  \citenamefont {Wang}, \citenamefont {Sun}, \citenamefont {Yang},
  \citenamefont {Comstock}, \citenamefont {Li}, \citenamefont {Zhou},
  \citenamefont {Zhang}, \citenamefont {You}, \citenamefont {Sun},\ and\
  \citenamefont {Liu}}]{JunLiu}%
  \BibitemOpen
  \bibfield  {author} {\bibinfo {author} {\bibfnamefont {K.}~\bibnamefont
  {Kim}}, \bibinfo {author} {\bibfnamefont {E.}~\bibnamefont {Vetter}},
  \bibinfo {author} {\bibfnamefont {L.}~\bibnamefont {Yan}}, \bibinfo {author}
  {\bibfnamefont {C.}~\bibnamefont {Yang}}, \bibinfo {author} {\bibfnamefont
  {Z.}~\bibnamefont {Wang}}, \bibinfo {author} {\bibfnamefont {R.}~\bibnamefont
  {Sun}}, \bibinfo {author} {\bibfnamefont {Y.}~\bibnamefont {Yang}}, \bibinfo
  {author} {\bibfnamefont {A.}~\bibnamefont {Comstock}}, \bibinfo {author}
  {\bibfnamefont {X.}~\bibnamefont {Li}}, \bibinfo {author} {\bibfnamefont
  {J.}~\bibnamefont {Zhou}}, \bibinfo {author} {\bibfnamefont {L.}~\bibnamefont
  {Zhang}}, \bibinfo {author} {\bibfnamefont {W.}~\bibnamefont {You}}, \bibinfo
  {author} {\bibfnamefont {D.}~\bibnamefont {Sun}},\ and\ \bibinfo {author}
  {\bibfnamefont {J.}~\bibnamefont {Liu}},\ }\bibfield  {title} {\bibinfo
  {title} {Observation of the chiral phonon activated spin seebeck effect},\
  }\href@noop {} {\bibfield  {journal} {\bibinfo  {journal} {XXX}\ }\textbf
  {\bibinfo {volume} {XX}},\ \bibinfo {pages} {xxxxxx} (\bibinfo {year}
  {2022})}\BibitemShut {NoStop}%
\bibitem [{\citenamefont {Mondal}\ \emph {et~al.}(2015)\citenamefont {Mondal},
  \citenamefont {Berritta}, \citenamefont {Carva},\ and\ \citenamefont
  {Oppeneer}}]{PhysRevB.91.174415}%
  \BibitemOpen
  \bibfield  {author} {\bibinfo {author} {\bibfnamefont {R.}~\bibnamefont
  {Mondal}}, \bibinfo {author} {\bibfnamefont {M.}~\bibnamefont {Berritta}},
  \bibinfo {author} {\bibfnamefont {K.}~\bibnamefont {Carva}},\ and\ \bibinfo
  {author} {\bibfnamefont {P.~M.}\ \bibnamefont {Oppeneer}},\ }\bibfield
  {title} {\bibinfo {title} {Ab initio investigation of light-induced
  relativistic spin-flip effects in magneto-optics},\ }\href
  {https://doi.org/10.1103/PhysRevB.91.174415} {\bibfield  {journal} {\bibinfo
  {journal} {Phys. Rev. B}\ }\textbf {\bibinfo {volume} {91}},\ \bibinfo
  {pages} {174415} (\bibinfo {year} {2015})}\BibitemShut {NoStop}%
\bibitem [{\citenamefont {She}\ \emph {et~al.}(2013)\citenamefont {She},
  \citenamefont {Fransson}, \citenamefont {Bishop},\ and\ \citenamefont
  {Balatsky}}]{PhysRevLett.110.026802}%
  \BibitemOpen
  \bibfield  {author} {\bibinfo {author} {\bibfnamefont {J.-H.}\ \bibnamefont
  {She}}, \bibinfo {author} {\bibfnamefont {J.}~\bibnamefont {Fransson}},
  \bibinfo {author} {\bibfnamefont {A.~R.}\ \bibnamefont {Bishop}},\ and\
  \bibinfo {author} {\bibfnamefont {A.~V.}\ \bibnamefont {Balatsky}},\
  }\bibfield  {title} {\bibinfo {title} {Inelastic electron tunneling
  spectroscopy for topological insulators},\ }\href
  {https://doi.org/10.1103/PhysRevLett.110.026802} {\bibfield  {journal}
  {\bibinfo  {journal} {Phys. Rev. Lett.}\ }\textbf {\bibinfo {volume} {110}},\
  \bibinfo {pages} {026802} (\bibinfo {year} {2013})}\BibitemShut {NoStop}%
\end{thebibliography}%

\end{document}